\documentclass{ws-procs975x65}

\begin{document}

\title{Conservative Formulations of General Relativistic Radiative Transfer\footnote{\uppercase{T}his work was supported 
by  \uppercase{S}cientific \uppercase{D}iscovery \uppercase{T}hrough
\uppercase{A}dvanced \uppercase{C}omputing (\uppercase{S}ci\uppercase{DAC}), a program of the \uppercase{O}ffice of \uppercase{S}cience of the \uppercase{U.S}. \uppercase{D}epartment of \uppercase{E}nergy (\uppercase{DoE}); and by \uppercase{O}ak \uppercase{R}idge \uppercase{N}ational \uppercase{L}aboratory, managed by \uppercase{UT-B}attelle, \uppercase{LLC}, for the \uppercase{DoE} under contract \uppercase{DE-AC05-00OR22725}.}}

\author{C. Y. CARDALL}

\address{Physics Division, Oak Ridge National Laboratory, 
Oak Ridge, TN 37831-6354, USA \\
Email: cardallcy@ornl.gov}

\address{Department of Physics and Astronomy, 
University of Tennessee, 
	Knoxville, TN 37996-1200, USA}%

\maketitle

\abstracts{
Accurate accounting of particle number
and 4-momentum in radiative transfer may be facilitated
by the use of transport equations that allow
transparent conversion between volume and surface integrals in both spacetime
and momentum space. Such conservative formulations of general relativistic
radiative transfer in multiple spatial dimensions are presented, and their
relevance to core-collapse supernova simulations described.}

\section{Neutrino transport in core-collapse supernovae}

Core-collapse supernovae result from the catastrophic collapse of the iron core of a massive star.  
Collapse is halted soon after the matter exceeds nuclear density. 
At this point a shock wave forms and begins to move out,
but soon stalls as energy 
is lost to neutrino emission and dissociation of heavy 
nuclei falling through the shock.
The details of how the stalled shock is revived
sufficiently to continue plowing through the outer layers of the
progenitor star are unclear. Some combination of neutrino heating of
material behind the shock, convection, instability of the spherical
accretion shock, rotation, and magnetic fields launches the explosion.

It is natural to consider neutrino heating as a mechanism for
shock revival. This is because neutrinos dominate the energetics of
the post-bounce evolution,
carrying away about 99\% of the released gravitational energy.
If we want to understand the explosion---which accounts for only about 1\% of 
the energy budget of the system---we should carefully account for the
neutrinos' much larger contribution to the energy budget, and accurately
track both the energy and lepton number transferred from the neutrinos to the fluid. 
In principle, this requires tracking both the energy and angle dependence of the neutrino distribution functions at every point in space. 
A full treatment of this six-dimensional neutrino radiation hydrodynamics
problem remains too costly for currently available computational resources;
the most that has been attempted so far has been a sort of ``3.5'' dimensional
approximation.\cite{buras03}

\section{The meaning of ``conservative''}

Conservative formulations of radiation transport are one aspect of trying to get the details of neutrino radiative transfer right. To give an idea of the kind of formulation
we seek, I first review ``elemental'' and ``conservative'' descriptions of the 
familiar dynamics of a fluid medium. 

The elemental formulation expresses the evolution of the fluid in terms of equations of motion for the 4-velocity spatial components $u^i$ and two quantities measured by a ÒcomovingÓ observer riding along with the fluid, the total energy density $\rho$ and baryon number density $n$: 
\begin{eqnarray}
(\rho +p){u^{\mu }}\Big(\frac{\partial {u^{i }}}{\partial {x^{\mu }}}+{{{{\Gamma }^{i }}}_{{\rho\mu }}}{u^{\rho }}\Big)+({g^{{i \mu }}}+{u^{i }}{u^{\mu }})\frac{\partial p}{\partial {x^{\mu }}}&=&0, \label{euler} \\
{u^{\mu }}\frac{\partial \rho }{\partial {x^{\mu }}}+\frac{(\rho +p)}{{\sqrt{-g}}}\frac{\partial }{\partial {x^{\mu }}}\big({\sqrt{-g}}{u^{\mu
}}\big)&=&0, \label{energy} \\
{u^{\mu }}\frac{\partial n}{\partial {x^{\mu }}}+\frac{n}{{\sqrt{-g}}}\frac{\partial }{\partial {x^{\mu }}}\big({\sqrt{-g}}{u^{\mu
}}\big)&=&0.\label{baryon}
\end{eqnarray}
The name ``elemental'' denotes the fact that by writing down separate equations of motion for the velocity and two comoving-frame
quantities, the kinetic and ``intrinsic'' fluid energies---two ``elements'' of the system---are analytically separated.

In contrast, the conservative approach expresses the evolution of the system in terms of the divergence of the stress-energy tensor $T^{\mu\nu}=(\rho+p)u^\mu u^\nu+p g^{\mu\nu}$ and the divergence of a baryon number flux vector $N^\mu=n u^\mu$:
\begin{eqnarray}
{\frac{1}{{\sqrt{-g}}}\frac{\partial }{\partial {x^{\mu }}}\big({\sqrt{-g}}{T^{\nu \mu }}}\big)&=&-{{{{\Gamma
}^{\nu }}}_{{\rho \mu }}}{T^{{\rho \mu}}},\label{divergence}\\
\frac{1}{{\sqrt{-g}}}\frac{\partial }{\partial {x^{\mu }}}\big({\sqrt{-g}}{N^{\mu }}\big)&=&0.\label{baryon2}
\end{eqnarray}
Volume integrals of the left-hand sides---obtained by multiplying by the invariant
spacetime volume element $\sqrt{-g}\,d^4x$ and integrating---are related to surface integrals in an obvious manner:
\begin{equation}
\int_V \sqrt{-g}\,d^4x\; {1\over\sqrt{-g}}{\partial\over\partial x^\mu}(\cdots^\mu)=\oint_S d^3 S_\mu (\cdots^\mu).
\end{equation}
Physically, this formulation relates the time rates of change of 4-momentum and baryon number in a volume to fluxes through a surface surrounding that volume; hence the labeling of this formulation as ``conservative.''

I will now describe
why a common equation of particle (or radiation) transport---the Boltzmann 
equation---is an ``elemental'' formulation.
The Boltzmann equation describes the evolution of $f$, the scalar distribution 
function: 
\begin{equation}
p^{\hat\mu}{\mathcal{L}^\mu}_{\hat\mu}{\partial f\over\partial x^\mu} -
{\Gamma^{\hat j}}_{\hat\mu\hat\nu}p^{\hat\mu}p^{\hat\nu}
{\partial u^{\hat i}\over\partial p^{\hat j}}{\partial f\over\partial 
u^{\hat i}} = \mathbb{C}[f]. \label{boltzmann}
\end{equation}
The scalar distribution $f$ is a function of time, spatial position, and momentum. Multiplied by suitable volume elements in phase space, it gives the number of particles in these infinitesimal cells.
The left-hand side of Eq. (\ref{boltzmann}) describes advection through phase space; it is closely related to the geodesic equations describing classical particle trajectories. The right-hand side, the collision integral $\mathbb{C}$, describes the change in the number of particles occupying a given trajectory.
Note that the spacetime derivative is taken with respect to coordinates $x^\mu$ with a plain, unaccented index. This represents a global coordinate basis; in the spatially multidimensional case it will typically be a ``lab frame'' basis. A global time coordinate is convenient for numerically integrating the system forward in time.
On the other hand, note the hats that accent the indices of the momenta. It is convenient to describe collisions in terms of momentum components measured with respect an orthonormal coordinate system carried by observers riding along with the fluid with which the neutrino radiation interacts. This ``comoving basis'' is indicated by the hatted indices. The tranformation ${\mathcal{L}^\mu}_{\hat\mu}$ relates the global coordinate basis to the comoving orthonormal basis, first through a transformation to a ``lab frame'' orthonormal basis, followed by a Lorentz boost to the comoving frame.
Notice also that the momentum coordinates have been changed from $p^{\hat i}$ to $u^{\hat j}$. This represents a change from Cartesian momentum coordinates to spherical momentum coordinates (energy and direction angles).
Finally, note the connection coefficients $\Gamma$. These represent  gravitational redshifts and trajectory bending induced by the curvature of spacetime, Doppler shifts and angular aberrations related to the Lorentz boost between the lab and comoving frames, and the use of curvilinear coordinate systems.

In terms of the categories described above in connection with fluid evolution, the Boltzmann equation is ``elemental'': The most fundamental quantity---the distribution function---is the evolved variable, and volume integrals of the equation in both spacetime and momentum space are not obviously related to surface integrals.

\section{The conservative formulations}

Having discussed what I mean by ``conservative,'' and having seen that the Boltzmann equation is ``elemental'' or non-conservative, I now turn to conservative formulations of radiation transport,\cite{cardall02} 
beginning with number conservation.
The first momentum moment of the scalar distribution function constitutes a particle number flux vector:
\begin{equation}
N^\mu = \int{\mathcal{L}^\mu}_{\hat\mu} p^{\hat\mu}\,f\,dP=\int\mathcal{N}^\mu\,dP, 
\label{numberVector} 
\end{equation}
where the quantity 
$\mathcal{N}^\mu={\mathcal{L}^\mu}_{\hat\mu}p^{\hat\mu} \,f$
is the contribution of each {\em comoving frame} momentum bin to the {\em lab frame} number vector. The particle number flux vector $N^\mu$ satisfies a law of particle number conservation, 
\begin{equation}
{1\over\sqrt{-g}}{\partial\over\partial x^\mu}\left(\sqrt{-g}\,N^\mu \right)
= \int\mathbb{C}[f]\,dP,\label{numberBalance}
\end{equation} 
which one would use to check number conservation.

Noting that the right-hand side of Eq. (\ref{numberBalance}) is simply the momentum integral of the collision integral, we see that this conservation law must simply be the momentum integral of Eq. (\ref{boltzmann}).
But it is not obvious how this works out, because of quantities  outside the spacetime and momentum 
derivatives that depend on
spacetime position and momentum.
Clearly, some conspiring cancellations are at work. But in the discretized case, in numerical work, a straightforward momentum integration of a straightforwardly discretized Eq. (\ref{boltzmann}) will not be consistent numerically with a straightfoward discretization of Eq. (\ref{numberBalance}).

However, the following expression is equivalent to the Boltzmann equation, yet transparently related to Eq. (\ref{numberBalance}): 
\begin{eqnarray}
{1\over\sqrt{-g}}{\partial\over\partial x^\mu}\left(\sqrt{-g}\,
\mathcal{N}^\mu\right) + & & 
\nonumber\\
\epsilon
\left|\det\left(d{\bf p}\over d{\bf u}\right)\right|^{-1} 
{\partial\over\partial u^{\hat i}}\left(-
{\Gamma^{\hat j}}_{\hat\mu\hat\nu}{p^{\hat\nu}{\mathcal{L}^{\hat\mu}}_\mu\over\epsilon}
\left|\det\left(d{\bf p}\over d{\bf u}\right)\right|
{\partial u^{\hat i}\over\partial p^{\hat j}}\, \mathcal{N}^\mu\right) &=& \mathbb{C}[f].
\label{numberConservative}
\end{eqnarray}
Multiplying Eq. (\ref{numberConservative}) by 
$dP=\epsilon^{-1}\left|\det\left(d{\bf p}\over d{\bf u}\right)\right|du^{\hat 1}du^{\hat 2}du^{\hat 3}$ and integrating immediately yields the number conservation law, Eq. (\ref{numberBalance}). Hence straightfoward differencings of these two expressions will be numerically consistent.

Another reformulation of the Boltzmann equation is an 
energy-conservative form, not detailed here. Suffice it to say that its momentum integral gives a transparent connection to the divergence of the particle stress-energy tensor.

The number- and energy-conservative formulations---which respectively facilitate
computation of lepton number and energy transfer to essentially machine 
accuracy---might be used in a couple of different ways. Both formulations could be solved separately, in order to nail down the transfer of both energy and lepton number. The values of the scalar distribution function implied by these two different distributions would then serve as a consistency check. Alternatively, only one of the formulations might be solved, with the analytic relationship between the two conservative formulations\cite{cardall02} being used to design finite difference expressions that provide consistency with the other formulation. This latter 
philosophy has been employed in the work of the group centered at 
Oak Ridge National Laboratory.\cite{liebendoerfer02}

\end{document}